\documentclass{article}

\def\giorno{13 December 2001}

\def \pn{\par\noindent}

\def\.#1{\dot #1}

\def\B{{\cal B}}
\def\F{{\cal F}}
\def\G{{\cal G}}
\def\I{{\cal I}}
\def\L{{\cal L}}

\def\R{{\bf R}}  

\def\X{{\cal X}}

\def\ss{\subset}
\def \pa{\partial}

\def\=#1{\bar #1}
\def\~#1{\widetilde #1}
\def\.#1{\dot #1}
\def\^#1{\widehat #1}

\def\xd{{\dot x}}
\def\yd{{\dot y}}
\def\vb{{\bf v}}
\def\wb{{\bf w}}
\def\grad{\nabla}     
\def\ker{{\rm Ker}}
\def\ran{{\rm Ran}}
\def\cd{\cdot}
\def\({\left(}
\def\){\right)}
\def\[{\left[}
\def\]{\right]}

\def\a{\alpha}
\def\b{\beta}

\def\de{\delta}   
\def\eps{\varepsilon}
\def\phi{\varphi}
\def\la{\lambda}
\def\s{\sigma}
\def\om{\omega}

\def \ep{\varepsilon}
\def \eps{\ep}

\def\phi{\varphi}
\def\Ga{\Gamma}

\begin{document}

\title{{\bf Poincar\'e normal and renormalized forms}}

\author{Giuseppe Gaeta\footnote{Paper to appear in {\it Acta Applicandae Mathematicae}; revised version. Work supported in part by ``Fondazione CARIPLO per la ricerca scientifica'' under project ``Teoria delle perturbazioni per sistemi con simmetria''.} \\ Dipartimento di Matematica, 
Universit\'a di Milano \\ V. Saldini 50, I--20133 Milano (Italy) \\ 
{\it gaeta@roma1.infn.it}}
 
\date{\giorno}
\maketitle
{\bf Summary.} We briefly review the main aspects of (Poincar\'e-Dulac) normal forms; we have a look at the non-uniqueness problem, and discuss one of the proposed ways to ``further reduce'' the normal forms. We also mention some convergence results. 

\section{Poincar\'e normal forms and their use}

Normal forms were introduced by Poincar\'e  as a tool to integrate nonlinear systems; by now we know this is in general impossible, but it turned out that the usefulness of normal forms goes well beyond integrability. 
 
An introduction to normal forms is provided e.g. in \cite{GMDE,EMSI,CGs,Gle,GH,IA,Ver}; see also \cite{Bel,Elp}. An introduction to normal forms for Hamiltonian systems is given in appendix 7 of \cite{MMCM}.

Their use is, beyond integrability issues, of three kinds:

\begin{itemize}
\item Classification of dynamical systems;
\item Qualitative study of dynamical systems;
\item Quantitative study of dynamical systems.
\end{itemize}

In this note I want to discuss the Poincar\'e-Dulac approach, and an extension thereof, focusing on the simpler setting, i.e. normal forms for a smooth dynamical system (equivalently, an autonomous first order ODE, a vector field) in $\R^n$, around an equilibrium point. For more general settings, see e.g. \cite{GMDE,EMSI}.  

I consider a first order ODE (dynamical system) in $R^n$ having a fixed point in the origin, and written as a series of homogeneous polynomial terms:
$$ \xd \ = \ A x \, + \, \sum_{k=1}^\infty f_k (x) \eqno(1) $$ 
where $f_k (a x) = a^{k+1} f_k (x)$ (we say also $f_k \in V_k$, and write $V = \oplus_{k=0}^\infty V_k$). We can of course think this is obtained from a general smooth\footnote{By this we will always mean $C^\infty$.} equation $ \xd = f (x)$, with $f(0)=0$, expanding it\footnote{In this way we lose all the information ``beyond all orders'' in perturbation theory, such as $e^{-1/x^2}$ terms; for this, see the works by Ecalle. Notice that if we deal with perturbative power series expansions of smooth dynamical systems (vector fields), it would be natural in some aspect to consider $C^\infty$ rather than analytic conjugacy.} around $x=0$. Actually, for most of our discussion, we could as well allow the series on the r.h.s. of (1) to be just a formal series (i.e. not require convergence in any neighbourhood of the origin).

I will also {\it assume that $A$ is semisimple}\footnote{Let me stress, however, that a good part of the discussion conducted here will also apply to the case where $A$ is not semisimple (at least provided it has nonzero semisimple part).}; for a discussion of the general case, see \cite{IA,Wal} and \cite{EMSI,Bru,Elp}. Recall that if $A$ is semisimple, it can be brought to diagonal form, and thus to be normal, $[A,A^+]=0$ (this will be the relevant property); it will thus be more economic to perform all computations in a basis where $A$ is diagonal. 
 
The idea of Poincar\'e was to eliminate nonlinear terms by a sequence of near-identity changes of coordinates,
$$ x \ = \ y \, + \, h_m (y) \eqno(2)$$
with $h_m \in V_m$, and $m$ taking successively the values $m=1,2,...$.
 
With this, and writing $D$ for the matrix of partial derivatives of $h_m$, eq.(1) reads $ [I+D] \yd  =  f ( y  + h_m (y) )$, which means
$$ \yd \ = \ [I+D]^{-1} \, \sum_{k=0}^\infty \, f_k (y + h_m (y) ) \ :=  \sum_{k=0}^\infty \, \~f_k (y) \eqno(3) $$
 
By a simple computation, we obtain that $\~f_k (x)
= f_k (x)$ for $k < m$, and (see below for the notation)
$$ \~f_m (x) \ = \ f_m (x) - \L_0 (h_m ) \ ; \eqno(4) $$ 
the effect on higher order terms is more involved, and is usually not considered in detail, as higher order terms can be normalized at later stages.
 
In eq.(4) above, $\L_0$ is  a linear operator, defined in terms of the linear part $A$ of $f$ as
$$ \L_0 (\psi ) \ := \ (Ax \cdot \grad ) \psi \, - \, A \psi \ ; \eqno(5) $$
notice that $\L_0 : V_k \to V_k$. Actually it is more convenient to describe this is terms of a general  bracket
$$ \{ \phi,\psi \} \ := \ (\phi \cdot \grad ) \psi \, - \,
(\psi \cdot \grad) \phi  \eqno(6) $$
(this is just the representation of the commutator of the vector fields $\phi^i \pa_i$ and $\psi^i \pa_i$ in terms of their components); with this, $\L_0$ is just defined as $ \L_0 (\psi) := \{Ax,\psi\}$.
 
In this way we can eliminate all the terms which are in the range of $\L_0$ (note that if $A$ is normal, $\ker (\L_0 )$ is
complementary to $\ran (\L_0)$ in $V$, and thus the whole normal form can be taken to be in $\ker (\L_0 )$; this means that the whole normal form commutes with the linear part of the system).
We can also introduce a scalar product\footnote{With the notation introduced below, the scalar product in $V$ can be chosen to be the Bargmann scalar product \cite{IA}, $(x^\mu \vb_r ,
x^\nu \vb_s ) := \delta_{r,s} \delta_{\mu , \nu} (\mu !)$. With this, $\L_0^+ (\psi) = \{ A^+ x , \psi  \}$.} in $V$, so that $\ran (\L_0) = \ker (\L_0^+)$.

We can now also easily characterize (with $A$ taken to diagonal form)  the terms in $\ker (\L_0^+)$; assume $\la_1 , ... , \la_n$ are the eigenvalues of $A$, and $\vb_1 , ... , \vb_n$ the corresponding eigenvectors (the computations are immediate if we take a basis of eigenvectors in $R^n$). Then a
vector $\wb = x^\mu \vb_r := x_1^{\mu_1} ... x_n^{\mu_n} \vb_r$ is in $\ker (\L_0^+)$ if and only if there are nonnegative integers $\mu_1 ... \mu_n$ (with $|\mu | := \sum_{s=1}^n \mu_s  > 1$) such that 
$$ \sum_{s=1}^n \mu_s \la_s \ = \ \la_r \eqno(7) $$
and $\ker (\L_0^+)$ is spanned by such ``resonant monomials''. The equation (7) is called a {\it resonance relation} of order $|\mu|$.
 
Thus, following the Poincar\'e  procedure, we can always
eliminate term by term all the nonlinear terms up to any desired order $N$, provided the $\la_i$ are nonresonant; note that this is generically the case.

In the case where the $\la_i$ satisfy some resonance relation, i.e. when (7) admits some solution, we can always eliminate nonresonant terms, and arrive at a form of $f$, the {\it Poincar\'e-Dulac normal form}, in which only resonant terms are present. That is, we arrive at coordinates in which (1) is written as\footnote{More precisely, we can arrive to such a form up to order $N$, for any desired $N$. That is, we should add a remainder ${\cal R}_N$ of arbitrarily high order $N+1$ in this equation.} 
$$ \xd \ = \ A x \, + \, \sum_{k=1}^\infty g_k (x) \ \ ; \ \ g_k \in \ker (\L_0^+ ) \ . \eqno(8) $$
It should be stressed that this result also holds for non-semisimple $A$ \cite{Elp}.

In this way we obtain a description of nonlinear systems (in normal form) ``compatible'' with a given linear part.
 
In practice, the normalization will be carried out
only up to some finite order $N$; one will then use perturbative arguments to infer that the dynamics of the full system is only a small modification -- at least in a sufficiently small neighbourhood of the origin (and/or for sufficiently short times) -- of that described by the truncated normal form.
In this way, if we were able to study the most general (truncated) system in the form (8) for given $A$, we would in principles obtain a classification of the qualitative local behaviours of nonlinear systems around a fixed point with prescribed linearization $A$.
 
Let me point out, after mentioning these ``qualitative'' uses of normal forms, the basis for its use in quantitative studies. Consider again (1), and denote by $\mu$ the first $k$ with $f_k \not\equiv 0$ (so generically $\mu = 1$); rescale $x$ as $x = \eps \~x$, divide by $\eps$, and drop the tilde for ease of notation, so that we have a system of the form
$$ \xd \ = \ A x \, + \, |\eps|^\mu g (x;\eps) \ ; \eqno(9) $$ 
let $ x(t) = \Phi_\eps (x_0 , t) $ be its solution with initial datum $x(0) = x_0$. If we approximate this with the
solution $\xi (t) = \Phi_0 (x_0 , t)$ to the system $\xd = Ax $ obtained for $\eps=0$, we are making an error
$$ \rho_\eps (t) \ := \ |x(t) - \xi (t) | \
= \ | \Phi_\eps (x_0 , t) \, - \, \Phi_0 (x_0 , t) | \eqno(10) $$
 
Assume that $A$ has norm $C$, and that in a region $\B$ we have $|g(x;\eps)| \le M (\eps)$ (all of our estimates will be only valid when $x$ and $\xi$ are both in $\B$, i.e. for a time $t$ depending on the exit time from $\B$). Then one obtains easily\footnote{Indeed $d \rho_\eps / dt \le | A (x - \xi) | + |\eps^\mu g
(x,\eps)| \le C \rho_\eps + |\eps|^\mu M (\eps) $.
Thus the growth of $\rho_\eps (t) >0$ is slower than that of a function $r(t)$ obeying $dr /dt = a r + b$ with $r(0)=0$ and $a = C$, $b= |\eps|^\mu M (\eps)$. The solution to this equation is $r(t) = (b/a) [\exp (at) - 1]$.} that  
$$ \rho_\eps (t) \ \le \ |\eps |^\mu {M \over C} \[ e^{Ct} - 1\] \ . \eqno(11) $$
This also means that $\rho_\eps (t) < \delta$ for a time not smaller than 
$$ t_0 \ = \ {1 \over C} \ \ln \[ 1 + {\delta C \over |\eps|^\mu M} \] \ . \eqno(12) $$ 
 
When we perform a normalization, with no resonance of order smaller than $\mu$, we are pushing nonlinearities to higher orders and thus increasing $\mu$ in the previous estimates; that is, we are improving the accuracy of the linear estimate, or extending the time interval over which we have an error of the same ``acceptable'' size $\delta$.
 
It should be stressed that in changing coordinates we are also changing the value of the constant $M$; if the ``small denominators'' appearing due to the change of coordinates are too small, their effect will more than counterbalance the gain we obtain by pushing up the nonlinearities, and disrupt the perturbative expansion.
 
\section{The problem of convergence}
 
It should be noted that the discussion conducted above is purely formal: the changes of coordinates are constructed as a series which could -- and in general will -- be only formal. 

Here we want to consider the case of analytic systems, and for these  we should study the convergence of the series defining the transformations. 
 
This is in general a hard problem (let us stress that it is difficult to prove convergence, but it is also difficult to prove divergence; see \cite{GMDE,EMSI}), and there are very few general results. Let me recall the main ones\footnote{It is remarkable that when we can guarantee convergence of the normalizing transformation on the basis of the first four criteria below, we have an {\it integrable normal form.}} \cite{GMDE,CGs,CGp,Wal}. Here the function $f: \R^n \to \R^n$ defining the dynamical system ${\dot x} = f(x)$ should be thought of as $f : {\bf C}^n \to {\bf C}^n$.

$\bullet$  {\it Poincar\'e criterion:} if the convex hull of the $\la_i \in {\bf C}$ does not include zero, then the normalizing series converges in an open neighbourhood of the origin.
  
$\bullet$  {\it Siegel criterion:} call the vector $\la = ( \la_1 , ... , \la_n ) $ of type $(C,\nu)$ if for each $r=1,...,n$ and for any $|\mu| \ge 2$  it satisfies $|\sum_s (\mu_s \la_s) - \la_r | \ge C/|\mu|^\nu $. If $f$ is analytic and $\la$ is of type $(C,\nu)$ for some positive numbers $C$ and $\nu$, then $f$ is analytically equivalent to its linear part. (Note that the set of vectors $\la$ which are not of type $(C,\nu)$ for any $C > 0$ has measure zero for $\nu > (n-2)/2$; in the real case this holds for $\nu > n-1$).

$\bullet$  {\it Pliss criterion:} Pliss realized that Siegel condition can be weakened: call the vector $\la = ( \la_1 , ... , \la_n ) $ of type $P(K,\nu)$ if for each $r=1,...,n$ and for any $|\mu| \ge K$, it satisfies either the resonance condition $|\sum_s (\mu_s \la_s) - \la_r | = 0$, or $|\sum_s (\mu_s \la_s) - \la_r | \ge 1/|\mu|^\nu $. 
 If $f$ is analytic, $\la$ is of type $P(K,\nu)$ for some positive numbers $K$ and $\nu$, and $f$ is formally conjugated to its linear part, then $f$ is analytically conjugated to its linear part.
 
$\bullet$  {\it Bruno criterion:} say that $\la$ satisfies ``condition  $\om$'' if $\sum_{k=1}^\infty 2^{-k} \ln (1/\om_k) < \infty $, where $\om_k$ is the minimum of $|\sum_s \mu_s \la_s - \la_r|$ on all the $r=1,...,n$ and all the $|\mu|<2^k$ not satisfying a resonance relation (condition $\om$ is weaker than the Siegel or Pliss condition \cite{Bru}). Say that $f$ having linear part $Ax$ satisfies ``condition A'' if\footnote{In geometrical terms, if $f$ is orbitally  equivalent to its linear part \cite{Wal2}. This means the trajectories of $f$ are the same as those of $Ax$.} $f(x) = [1 + \a (x)] A x$. Then, if $\la$ satisfies
condition $\om$ and $f$ is formally conjugated to a normal form $\^f$
satisfying condition A, then $f$ is conjugated to $\^f$ by an analytic transformation. (Condition A can be weakened for suitable classes of $f$'s).
  
$\bullet$  {\it BMW-C theory:} assuming condition $\om$ for $\la$, condition $A$ can be replaced by suitable symmetry conditions on $f$. This theory was created by Markhashov \cite{Mar} and by Bruno and Walcher \cite{BW}; it was later extended by Cicogna \cite{Cic} and again by Walcher \cite{Wal4}. This matter is discussed in \cite{CGp}
and in the SPT2001 tutorial by Cicogna and Walcher \cite{CWt}. See also the recent work by Stolovich \cite{St1} which, together with many other things, contains results in this direction. 
  
$\bullet$  {\it Normalization of an algebra of vector fields:} one can 
consider the problem of simultaneously normalizing an algebra of (analytic) vector fields $X_i$ around a common singular point. Vey \cite{Vey} proved that if they span an abelian algebra and their linearizations at the singular point are linearly independent, then they are simultaneously and analytically normalizable. Results in this direction are also contained in \cite{ABP,GY,St2}; see also \cite{CGs}.

$\bullet$  {\it Hamiltonian systems:} in the case of Hamiltonian systems, we can deal directly with the Hamiltonian (i.e. a scalar function) rather than with the hamiltonian vector field this generates. Many of the results quoted above can be recast in this setting, but there have been investigations of the convergence in the hamiltonian case {\it per se}. In this context, we refer to
the work of Russmann, Bruno, Vey, and Ito \cite{Bru,Ito,Rus,Vey}, as well as to that of Giorgilli and collaborators \cite{Gio1} (also in connection  with the study of effective stability and KAM theory \cite{Gio2,Gio3}). I will not discuss these results here; see however the paper by Cicogna in \cite{SPT1} and \cite{CGs,CGp} for brief discussions.

$\bullet$  {\it BCS theory:} convergence of the normalizing transformation implies that the original vector field and its normal form are analytically conjugated. In many cases one would be satisfied with $C^\infty$ conjugation \cite{Bel,Be3} (or even with $C^k$ conjugation: e.g., this is the only one making sense for $C^k$ vector fields. For $C^k$ normal forms theory, see \cite{IY}). It was proved by Chen and by Sternberg \cite{Che,Ste} that if the linear part $A$ of the system is hyperbolic, then formal conjugacy implies $C^\infty$ conjugacy (see also \cite{Bel,Dem,Har}). This result has been recently extended to equivariant (or reversible) systems, and also to the Hamiltonian framework, by Belitskii and Kopanskii \cite{BeK}.     

$\bullet$  {\it Orbital conjugacy:} the reader should be warned that in many cases one is satisfied with orbital equivalence (two systems are orbitally equivalent if they have the same integral curves, independently of the time law they follow on these). This question is considered e.g. in \cite{EMSI,Ily,Wal2}.

\medskip

In the following, I will come back to considering only formal
transformations; for a discussion of convergence see e.g.
\cite{CGs,CGp,CWt}.

\section{Some remarks on Poincar\'e normalization}

Now that I described the why and how of Poincar\'e normalization, let me remark some simple points which are relevant for the following.
  
\begin{itemize}

\item The change of coordinates needed to normalize $f$ at order $k$ is generated by $h_k$ solving $\L_0 (h_k) = \pi_0 (f_k)$, where $\pi_0$ is the projection from $V$ to the range of $\L_0$, and $f_k$ is the term obtained after the first $(k-1)$ normalizations. The solution to this ``homological equation'' for $h_k$ is unique up to a term in $\ker (\L_0)$. That is, if $\de h_k = h_k - \^h_k \in \ker (\L_0)$, then $h_k$ and $\^h_k$ define the same transformation in $V_k$.
  
\item On the other hand, functions $h_k$ and $\^h_k$ as above have in general a different effect on the $f_p$ with $p>k$. Thus the freedom in the choice of $h_k$ (for resonant systems) means that we can generate different normal forms starting from the same nonlinear system.
  
\item This shows at once that normal forms are somehow ambiguous, i.e. the normal forms classification is in general redundant: different normal forms (with the same linear part) can be actually conjugated.
  
\item It is maybe worth remarking explicitely that the higher order terms generated by the change of coordinates $x = y + h_k (y)$ will in general also have a resonant part; this cannot be eliminated by the Poincar\'e procedure at later stages\footnote{Thus, in particular for resonant systems, it is possibly misleading to say that normalization is a way to ``simplify'' (a rather vague word) nonlinear systems: a system with very few terms can have infinitely many terms once taken into normal form.}.
  
\end{itemize}

It is clear that from the point of view of qualitative study and
classification we would like to eliminate or at least reduce the redundancy in the normal forms classification. This would also have an obvious computational advantage: among normal forms corresponding to $f$ we could choose the most convenient for our needs.
  
Thus the problem of ``further normalization'' has been considered by several authors\footnote{Actually already Dulac (see page 335 of \cite{Dul}) remarked that possibly a choice of the terms $\de h_k$ in $\ker (\L_0 )$ different from the one he was employing -- i.e. $\delta h_k = 0$ -- could have produced a simpler normal form.}; I will quote first of all Takens \cite{Tak}, and Broer who in his thesis (advised by Takens) set the problem -- and the theory of normal forms -- in terms of filtrations of Lie algebras \cite{Bro1,Bro2,BrT}. In this respect, one should also mention the early work of Kummer \cite{Kum}. 

Broer's Lie algebra approach was also followed by van der Meer \cite{vdM} and by Baider, Churchill and Sanders  \cite{Bai,BaC,BaS}; in particular they obtained a formulation of ``unique normal forms'', see also \cite{ChK,KOW}. Unfortunately, their results are very difficult to implement on concrete systems, even the simpler ones, and seems to be not applicable in concrete terms (on the other hand, Broer's approach is a guiding light for many questions in normal forms theory).
  
Other, and more recent, attempts in this direction (some of them with
different tools) include works by Algaba, Freire and Gamero, by Chen and Della Dora, by Kokubu, Oka and Wang, by Ushiki, and by myself \cite{AFG,CDD,KOW,Gae,Ush}.

Unfortunately, no systematic exposition and comparison of these different approaches is available; needless to say, this is not the place to attempt such a work, but maybe some reader of the present paper will endeavour a comprehensive discussion of this matter, which would surely be useful.

The same problem of further normalization was also tackled, of course, by russian authors. One should quote in this respect at least the works by Belitskii \cite{BelNF} and by Bruno \cite{Bru}; these are also discussed and compared in the recent book \cite{Bru2}. For russian work in the area see also, of course, \cite{EMSI}.

In the next sections I will present a simple way\footnote{The reader should be warned that the discussion of it given in \cite{Bru2} uses  different basic definitions; see \cite{GaB}.} to (in general, only partially) remove the degeneration of normal forms. This is based on ideas which I believe were first introduced by Broer (and recalled in a somewhere deformed way in the final section 6); the approach discussed below in sections 4 and 5 aims at a smaller generality than Broer's one, but is easily implemented in practice \cite{Gae,GaB}. 

I would like to thank prof. Belitskii for pointing out, after a preliminary version of the present paper was circulated, that this approach is very much similar to the one proposed in his work \cite{Bel5}.

\section{Renormalization (further normalization)}
 
\def\LP{Lie-Poincar\'e }
\def\H{{\cal H}}
\def\M{{\cal M}}
\def\ran{{\rm Ran}}
 
The approach I will discuss in this section (based on \cite{Gae}; see also \cite{CGs}, and \cite{GaB} for more recent discussions) is completely algorithmic, and not any more difficult to implement (by hand or by computer) than the standard Poincar\'e one: actually it is just based on a careful exploitation of the freedom left by Poincar\'e algorithm. As mentioned above, this is quite similar to the one proposed in \cite{Bel5}.
  
The main idea is to control the effects of the change of variables generated by $h_k$ on higher order terms $f_p$, $p>k$; as already mentioned, this change of variables will in general produce resonant terms. This should not be seen as a drawback, but rather as an advantage! By a judicous choice of $h_k$, we can use this feature to {\it eliminate higher order resonant terms}. This ``judicious choice'' amounts to choosing the $h_k$'s as solutions to ``higher homological equations''.
 
It is actually convenient to consider an improved version of Poincar\'e transformations, i.e. \LP  ones; for some of the advantages of using these see \cite{BGG}, see also \cite{Dep,Walx}.  The idea is to define a vector field 
$$ H \ := \ h_m^i (x) \, {\pa \over \pa x^i} \eqno(13) $$
in $\R^n$ and consider the flow $\Phi (\s ; x_0 )$ generated by this.  The change of coordinates is then given by $x = \Phi (1;y)$, which coincides with the standard one at lowest nontrivial order.

The change of coordinates with generator $h_m$ maps (1) into a new system $ \yd = \^f (y) = Ay + \sum_{k=1}^\infty \^f_k (y) $, where $\^f_k (x) = f_k (x)$ for $k < m$, while for $k \ge m$ we have, with $[a]$ the integer part of $a$,
$$ \^f_k (x) \ = \ \sum_{s=0}^{[k/m]} \, {1 \over s !} \, \H_m^s
(f_{k-sm}) \ ; \eqno(14) $$ 
here the operator $\H_m$ is defined as $\H_m (\a ) := \{h_m,\a \}$, and $\H_m^p$ means applying $\H_m$ for $p$ times. Note that the expression (14) is just the translation to the present notation of the familiar Baker-Campbell-Hausdorff formula: indeed it states
that the vector field $X = f^ i \pa_i$ is changed into a vector field $\^X = \^f \^\pa_i$ given by $ \^X := \[ e^{\s H} X  e^{- \s H}  \]_{\s =1} $.

Having determined the effect of the \LP  transformation to all orders, we can try to control it.
  
Let us consider again our system (1), i.e. $ \xd = f(x) = Ax +
\sum_{k=1}^\infty f_k (x)$.
We can normalize the quadratic term $f_1$ by solving the homological equation $ \L_0 (h_1)  =  \pi_0  f_1 $, and we get\footnote{The notation $\^f_1$ points out that this term is stable, i.e. the algorithm will not change it any more.} 
$$ \^f_1 \equiv f_1^{(1)} = f_1 - \L_0 (h_1 ) \ . \eqno(15) $$
 
The higher order terms $f_k$ ($k \ge 2$) will of course also change with this first change of variables, but for the sake of notation we will not introduce indices to stress this fact, and call the new expressions of these again $f_2 , f_3 , .....$

Now, consider the cubic term $f_2$; we normalize it by solving $\L_0 (h_2) = \pi_0 f_2$, and we get $ f_2^{(1)} = f_2 - \L_0 (h_2) $. If this is nonzero we would like to (attempt to) further eliminate some
terms. We can do it by exploiting the higher order effects of a transformation with generator $h_1^{(1)}$, and in order to avoid this affects  $f_1^{(1)}$ we have to require $h_1^{(1)} \in \ker (\L_0)$. From the general formula (14) we have then that with such a transformation 
$$ \^f_2 \ \equiv \ f_2^{(2)} \ = \ f_2^{(1)} \, + \, \{ h_1^{(1)} , f_1^{(1)} \} \ . \eqno(16) $$
 
We define the linear ``higher homological operators'' $\L_k$ as 
$$ \L_k (\a ) \ := \{ \^f_k , \a \} \eqno(17) $$ 
so that the above reads
$$ f_2^{(2)} \ = \ f_2^{(1)} \, - \, \L_1 \( h_1^{(1)} \) \ , \eqno(18) $$
and we can eliminate from $f_2^{(1)}$ all terms which are in the range of $\M_1$, defined as the restriction of $\L_1$ to $\ker (\L_0)$. Thus, if we choose the complementary spaces to the ranges considered as the orthogonal complement, we get $\^f_2 \in [\ran (\L_0)]^\perp \cap [\ran (\M_1)]^\perp$. 

To avoid any misunderstanding, we stress immediately that our procedure is based on these restricted operators $\M_k$, see below, and {\it not} on the full higher homological operators $\L_k$.
 
Higher order terms, $f_3$ and so on, will also change; again we will 
denote them, for the sake of notation, still as $f_3$ etc. not to introduce heavy notation.

Let us, for the sake of clarity, go one step further and consider the quartic term $f_3$. First we normalize it by a transformation generated by $h_3$, obtaining a $f_3^{(1)} \in \ker (\L_0^+)$.
Then we can still use a transformation generated by $h_2^{(1)} \in \ker (\L_0 )$ to eliminate some terms without affecting $f_2$; we obtain in this way a $f_3^{(2)}$. But we can also still use a transformation generated by $h_1^{(3)}$ provided this does not affect $\^f_1$ nor $\^f_2$, i.e. provided $h_1^{(3)} \in \ker (\L_0) \cap \ker (\L_1)$; with this, we obtain 
$$ \^f_3 \ = \ f_3^{(2)} \, - \, \L_2 \( h_1^{(3)} \) \ , \eqno(19) $$
i.e. we can eliminate also all the terms which are in the range of $\M_2$, the restriction of $\L_2$ to $\ker (\L_0) \cap \ker (\L_1)$.
 
At this point the general procedure should be quite clear; in order to state precisely the result I should define some chains of spaces and
operators\footnote{Note that all of these spaces could be intersected with, and operators restricted to, the $V_k$, obtaining finite dimensional spaces and operators \cite{Gae}. I will not do this to avoid a cumbersome notation, but in computer algebra implementation of this algorithm one would proceed in this way.}.
\medskip
 
\pn $\bullet$ Let $H^{(0)} := V$, and define $H^{(p+1)} := \ker (\L_0 ) \cap ... \cap \ker (\L_p)$. Notice that $H^{(p+1)} \subseteq H^{(p)}$.
 
\pn $\bullet$ Define the operators $\M_p  := \L_p \vert_{H^{(p)}}$.
 
\pn $\bullet$ Let $F^{(0)} := V$, and define $F^{(p+1)} := [\ran (\M_0 )]^\perp \cap ... \cap [\ran (\M_p)]^\perp$. Notice that $F^{(p+1)} \subseteq F^{(p)}$.

\medskip 
 
It will also be convenient to set a notation for relevant projection
operators:
 
\medskip 

\pn $\bullet$ Denote by $\chi_s$ the projection operator $\chi_s : V \to \ker (\L_s)$
 
\pn $\bullet$ Denote by $\mu_s$ the composition of projection operators $\mu_s = \chi_{s-1} \cd ... \cd \chi_0$ for $s\ge 1$; we set $\mu_0 \equiv I$.
 
\pn $\bullet$ Denote by $\pi_s$ the projection operator $\pi_s : V \to \ran (\M_s ) $.
 
\medskip 

Note that with these definitions, $H^{(p)} = \mu_p V$, and $\M_p = \L_p \circ \mu_p$.
 
\medskip

\pn{\bf Definition.} The dynamical system $\xd = f(x) = \sum_{k=1}^\infty f_k (x)$ is in {\it Poincar\'e renormalized form} (PRF) up to order $N$ if $f_k \in F^{(k)}$ for all $k < N$. In the formal limit $N \to \infty$, we say the system is in PRF, tout court.

\medskip

\pn{\bf Theorem.} {\it Any analytic or formal smooth dynamical system can be taken into PRF up to any desired order $N$ by means of a sequence of \LP  transformations.}

\medskip

\pn{\bf Proof.} Proceeding as above, the transformation generated by $h_{k-p}^{(p)}$ will take $f_k^{(p)}$ into 
$$ f_k^{(p+1)} \ = \ f_k^{(p)} \, - \, \M_p \( h_{k-p}^{(p)} \)
\eqno(20) $$
If we choose $h_{k-p}^{(p)}$ as a (non-unique) solution to the higher
homological equation 
$$ \pi_p \, f_k^{(p)} \, - \, \M_p \(h_{k-p}^{(p)} \) \
= \ 0 \eqno(21) $$
i.e. as 
$$ h_{k-p}^{(p)} \ = \ \mu_p \cd \M_p^{*} \cd \pi_p \[ f_k^{(p)} \] \eqno(22)  $$ 
where $\M_p^*$ is the pseudo-inverse to $\M_p$, then we arrive -- after applying the full procedure to $f_1 , ... , f_N$ -- to a system $ \xd = f^* (x) = \sum_{k=0}^\infty f_k^* (x) $ where $ f_k^* (x) = f_k^{(k)} \in F^{(k)} \cap
V_k$ for all $k \le N$. \hfill $\triangle$

\medskip

\pn{\bf Remark.} The proof is completely constructive, as (22) tells which generator should be chosen at each step (and this is of course the nontrivial content of the theorem). It can thus be transformed in an algorithm to be implemented on a computer. $\odot$ 
  
\pn{\bf Remark.} The algorithm can deal with systems having zero linear part, contrary to the standard normalization algorithm. For such an application in a symmetry context, see \cite{GaG}. $\odot$

\pn{\bf Remark.} The algorithm is only based on the grading of resonant terms (resonant vector fields) by their homogeneity properties, and is therefore completely general. However, it fails to recognize the Lie algebraic properties of the set of resonant vector fields. Not surprisingly, taking these into account leads to a more powerful reduction, which we discuss below. $\odot$

Notice that this procedure, as discussed here, is completely formal. A
discussion (certainly not final) of convergence of PRFs -- and also of PRF for Hamiltonian systems, and of PRF in the presence of symmetry -- is contained in \cite{CGs} and in my contributions to the SPT workshops \cite{SPT1,SPT2}. A discussion of these matters is certainly incompatible with the limited scope of this paper.

I would now like to briefly mention an example of application of PRFs to a concrete problem, i.e. the (formal) analysis at singular points of vector fields in the plane having purely rotational linear part; this case corresponds to the vector field being, in the first order approximation, an integrable hamiltonian vector field. A complete analysis of regular singular points of vector fields in the plane from the point of view of PRFs is contained in \cite{GaB}, where analytic computations are described in detail, and completely explicit computations are pursued up to order six (order ten in a special case).
\bigskip

\pn {\bf Example.} Let the dynamical system $\xd = f (x)$ in $\R^2$ have a singularity at the point $x_0$; we can always shift coordinates so that $x_0 = 0$. Let the linearization $\xd = [(\pa_j f^i)(x_0)] x^j \equiv A^i_j x^j$ of $X$ at $x_0 = 0$ be given by 
$$ A \ = \ \pmatrix{0 & -1 \cr 1 & 0 \cr} \eqno(23)$$
Then the standard normal form is given (in vector notation) by 
$$ \xd \ = \ A x \ + \ \sum_{k=1}^\infty \, r^{2k} \, (a_k I \, + \, b_k A ) \, x \eqno(24) $$
where $I$ is the identity matrix and $r^2 = |x|^2 = (x_1^2 + x_2^2)$. 

In general, $a_1,b_1$ are nonzero (but if the system is hamiltonian, all the $a_k$ vanish; this case is treated in \cite{FoM,SiM}); however, for the sake of generality, let us assume $a_k = 0$ for $k < \mu$ and $b_k = 0 $ for $k < \nu$, with $a_\mu \not= 0$ and $b_\nu \not= 0$.

Then, following step by step the procedure described above, the PRF is given by 
$$ \cases{
{\dot w} \ = \ A w \ + \ (r^{2 \mu} \a_\mu + r^{4 \mu} \a_{2 \mu} ) \, w & for $\mu < \nu$ \cr
{\dot w} \ = \ (1 + r^{2 \mu} \b_\mu ) \, A w \ + \ (r^{2 \mu} \a_\mu + r^{4 \mu} \a_{2 \mu} ) \, w & for $\mu = \nu$ \cr
{\dot w} \ = \ (1 + r^{2 \nu} \b_\nu ) \, A w \ + \ \sum_{k=\mu}^\infty (r^{2 k} \a_k ) \, w & for $\mu > \nu$ \cr}
 \eqno(25)$$
where $\a_k , \b_k$ are real numbers (in general different from the
$a_k, b_k$).

This result was obtained in \cite{Gae}; however the computation (and the result reported) there is not correct in the case with $\mu > \nu$; the correct computation leading to (25) is given in \cite{GaB}.

Notice that the standard NF is characterized by two infinite sequences of constants; the PRF is finite except in case $\mu > \nu$, where it however depends on only one infinite sequence of constants.

\section{Lie renormalization}

Let us look more closely at the set $\G$ of nonlinear vector fields resonant with a given linear one, $X_A = (Ax)^i \pa_i$. 

First of all, it is clear that they form a Lie algebra (in general infinite dimensional) under commutation. Actually this also belongs to   a finitely generated module, as we now briefly discuss recalling a general characterization of vector fields in normal form \cite{CGs,Elp,IA,Wal}. 

We say that the differentiable function $\phi : \R^n \to \R$ is an invariant for $X_A$ if $X_A (\phi ) = 0$. Denote by $\I (A)$ the set of invariants for $X_A$ which are meromorphic (that is, can be expressed as a quotient of analytic functions) in the $x$ coordinates; and denote by $\I_* (A) \ss \I (A)$ the set of analytic invariants for $X_A$.

Let $G = C(A)$ be the centralizer of $A$ in the algebra of $n$ dimensional matrices; let its Lie algebra be spanned by matrices $\{ K_1 , ... , K_d \}$ (we can always assume $K_1 = I$, and take one of the $K_\a$ to be $A$ if this is nonzero; notice that $d \le n$). We denote by $X^{(\a)}$ the vector fields corresponding to these, i.e. given in the $x$ coordinates by $X^{(\a)} = (K_\a x)^i \pa_i$. 

Then it results \cite{CGs,Elp,IA,Wal} that the most general vector field $W$ in $\G$ can be written as
$$ W \ = \ \sum_{\a=1}^d \ \mu_\a (x) \ X^{(\a)}  \eqno(26)$$
where $\mu_\a (x) \in \I (A)$; this means that $\G$ is contained in the module over $\I (A)$ generated by $G$. 

Notice that the vector field $W$ must be analytic in the $x$, while generic vector fields of the form $\mu_\a (x) X^{(\a)}$ are not such:  $\G$ is not the full $G$-generated module over $\I (A)$. 

In several cases it happens that $\G$ has a more convenient structure, i.e. the $\mu_\a$ in (26) can actually be taken to be in $\I_* (A)$, and not just in $\I (A)$. In this case we have $\G = \I_* (A) \otimes G$, and we say that all the vector fields in $\G$ are {\it quasi-linear}, or that we have a {\bf quasi-linear normal form}. In particular, this is e.g. the case when $A$ admits only one basic invariant. 

Call $\X_\a$ the algebra spanned by vectors which are written as $X = s (x) X^{(\a)}$ with $s \in \I (A)$; call $\X_\a^* \ss \X_\a$ the algebra spanned by vectors as above with $s \in \I_* (A)$ (this is the full module over $\I_* (A)$ generated by $X^{(\a)}$). 
As seen before, in general we have $ \X_1^* \oplus ... \oplus \X_d^*  \subseteq \G \subset \ \X_1 \oplus ... \oplus \X_d$, and in the quasi-linear case we actually have $\G = \X_1^*  \oplus ... \oplus \X_d^*$.

Consider now the commutation relations between elements $\mu_\a (\Psi ) X^{(\a)} $ and $\s_\b (\Psi ) X^{(\b)}$ of the subalgebras $\X_\a$ and $\X_\b$ respectively ; it is immediate to check that, with $\grad_i := \pa / \pa \psi_i$, we have 
$$ \begin{array}{l}
\[ \mu_\a X^{(\a)} , \s_\b X^{(\b)} \] \ = \ \( \mu_\a \, (\grad_i  \s_\b) \, X^{(\a)} (\psi_i) \) X^{(\b)} \ + \\
\ \ \ \ \ - \ \( \s_\b \, (\grad_i \mu_\a) \, X^{(\b)} (\psi_i) \) X^{(\a)} \ + \ \( \mu_a \s_\b \) \, \[ X^{(\a )} , X^{(\b)} \] \ . \end{array} \eqno(27)$$

\pn{\bf Remark.} Notice that when $X^{(\b)} = X_A$, by definition $X^{(\b)} (\psi_i ) = 0$, and $[X^{(\a)} , X^{(\b)} ] = 0$; thus the corresponding subalgebra $\X_\b$ is always an abelian ideal in $\G$. 
$\odot$

\pn {\bf Remark.} Note also that, as obvious from the formula (27) above, the direct sum of subalgebras $\X_{\a_1} \oplus ... \oplus \X_{\a_s}$ is a subalgebra in $\G$ if and only if $\{ X^{(\a_1 )} , ... , X^{(\a_s )} \}$ span a subalgebra in $G$. $\odot$

Assume now that $\G$ is quasi-linear, and that we are able to determine a sequence of subalgebras $\F_p \subseteq \G$, each of them being the direct sum of $\X_\a$ subalgebras, such that $\F_0 = \G$ and 
$$ \[ \, \G \, , \, \F_p \, \] \ = \  \F_{p+1} \ ; \eqno(28) $$
if this terminates in zero we say that $\G$ has a {\it quasi-nilpotent} structure. Notice that the factor algebras $\Gamma_p := \F_p / \F_{p+1}$ are in general {\it not } abelian.

By the above remark, $\G$ can have a (nontrivial) quasi-nilpotent structure only if $G$ is nilpotent. The chain of subalgebras $\F_p \subset \G$ can then be read off the descending central series $G_p$ of $G$; recall that the factor algebras $\gamma_p = G_p / G_{p+1}$ for this are abelian. The nonabelian subalgebras $\Ga_p$ introduced above are therefore modules over $\I (A)$ generated by abelian subalgebras $\gamma_p$ of $G$.

Assume now $\G$ is quasi-nilpotent. In this case we can first work with generators in $\Ga_1$ and simplify terms in $\Ga_1$ (e.g. by following the PRF algorithm within the set $\Ga_1$; this allows to work with more familiar projection and homological equations than if setting the problem in a completely Lie algebraic framework), then consider generators in $\Ga_2$ and simplify the corresponding terms being guaranteed that $\Ga_1$ terms are not changed, and so on. 

We will call the reduced normal form obtained in this way the {\it Lie renormalized form} (LRF); the Lie renormalization is thus a blend of Lie algebraic reduction along the chain of subalgebras identified by (28) and of PRF reduction in each of the $\Ga_\a$ subsets.

\pn{\bf Remark.} In this case we are -- roughly speaking --  just using the nilpotent structure of (the finite dimensional group) $G$, rather than the one of (the infinite dimensional algebra) $\G$. $\odot$

\pn{\bf Remark.} Needless to say, this approach is particularly convenient when the $\Ga_p$ are generated by a single element of $G$; it can however be applied also in other situations. $\odot$

Let us briefly discuss why we have considered the quasi-linear case rather than the general one. It should indeed be clear that the key feature is the assumption that we are able to identify $\F_p$ satisfying (28), so that this kind of approach is not confined to the quasi-linear case. However, recall that we have to consider smooth  vector fields as generators of our Lie-Poincar\'e changes of coordinates, and that at each step we have to deal with smooth  (partially renormalized) vector fields. As these are only a subset of $\X_1 \oplus ... \oplus \X_d$ (and one identified by analytic, not geometric, conditions), checking these conditions at each step will in general be a prohibitive task. The problem is not present for quasi-linear normal forms, when we deal with the full module $\I_* (A) \otimes G$.

Thus, in practice, the LRF approach is not always applicable; however, when it can be applied the computations required are particularly simple \cite{GaB}. We also stress that the LRF (like the PRF) for a given system is in general not unique.

We stress that the quasi-nilpotent structure is met in applications: e.g., it applies to any nontrivial two-dimensional case and several three-dimensional ones \cite{GaB}. More generally, it always (but not only) applies when there is only one basic invariant.
We consider now again a simple example; for other explicit ones see \cite{GaB}.
\bigskip

\pn {\bf Example.} Let us consider again the case considered at the end of the previous section. Proceeding as discussed in this section, we obtain the LRF
$$ {\dot w} \ = \ A w \ + \ (r^{2 \mu} \a_\mu + r^{4 \mu} \a_{2 \mu} ) \, w \ + \ \( \sum_{k=\nu}^\mu r^{2 k} \b_k \) A w $$
where the sum should be considered as zero for $\mu < \nu$.
Notice that now we get a finite dimensional reduced normal form in all cases.

\section{Filtering of Lie algebras}

I will finally briefly recall (and reformulate, hopefully without distorting it to such an extent to make it completely different from the original version) the general approach employing the Lie algebraic structure of $\G$ (recall this included nonlinear fields only) proposed by Broer and Takens \cite{Bro2,BrT}, and further developed by other authors.

The Broer procedure is always applicable, and is able produce a unique reduced normal form \cite{Bai,BaC}. Unfortunately, implementing it in practice seems to be quite hard, and this approach has been applied only to systems of very low dimension.
On the other side, the ideas put forward by Broer are at the basis of many approaches -- including the one depicted in previous sections -- to the problem of further reduction of normal forms.

One can consider in general terms the descending central series of $\G$ \cite{Kir,NaS}. We recall this is defined by $\G_0 = \G$ and\footnote{Needless to say, $\G_k$ is {\it not } the set of fields in $\G$ homogeneous of some degree in the $x$!} $\G_{k+1} = [ \G , \G_k ]$; as well known the factor algebras $\Ga_k = \G_k / \G_{k+1}$ are abelian. It is then possible to eliminate terms by inner automorphisms of $\G$ (that is, by acting on $\G$ with vector fields in $\G$) proceeding along $\G_k$, i.e. filtering the Lie algebra $\G$. 
This is relevant to our problem since Lie-Poincar\'e transformations always reduce to inner operations on $\G$.

This approach is, of course, particularly convenient when $\G$ is nilpotent; actually if $\G_k$ is the descending central series for $\G$, we have $\G_k \cap V_m = \emptyset$ for $m<k$. We can reduce to the more familiar case of algebras nilpotent of finite order $N$ by considering only the projections of $\G_k$ and $\Ga_k$ to the linear space $V^N = V_1 \oplus ... \oplus V_N \subset V$.

Let us now briefly describe the implementation of the Broer renormalization procedure in the language employed in the present paper. We consider the sequence of abelian factor algebras $\Ga_k = \G_k / \G_{k+1}$, and denote by $\chi_k$ the projection from $\G$ to $\Ga_k$. 

We will produce a sequence of vector fields $W^{(j)}_*$, each being the result of the first $j$ steps of further normalization. Let $\eta_j$ be the operator on $\G$ defined by $\eta_j (H) = e^H W^{(j)}_* e^{-H}$; let $\pi_j$ be the operator of projection from $\G$ to the range of $\eta_j$. We can reduce the normal form $W = W^{(0)}_*$ as follows. 

As the first step, consider a $H^{(0)} \in \G$ and require that $ \~W = W^{(1)}_* \ := \eta_0 (H)$ is such that $ \pi_0 [ \chi_0 \eta_0 (H^{(0)}) ] = 0 $. This determines (non uniquely) $H^{(0)}$, and produces a $W^{(1)}_*$. 

At further steps, we have the same setting; the ``homological equations'' on Lie algebras to be solved at each step will be 
$$ \pi_j \, \[ \, \chi_j \( \eta_j (H^{(j)}) \) \] \ = \ 0 \ ; \eqno(29)$$
this determines $H^{(j)}$. Each $W^{(j+1)}_*$ is then determined as 
$$ W^{(j+1)}_* \ = \ \eta_j (H^{(j)} ) \ := \ \exp [H^{(j)}] \, \exp [W^{(j)}_* ] \, \exp [ - H^{(j)} ] \ . \eqno(30) $$

\section*{Acknowledgements}

The first version of this paper was written while I was a guest of the Theory Group in the Dipartimento di Fisica, Universit\`a di Roma; I would like to thank them, and in particular Toni Degasperis, for their continuing and always kind hospitality. The paper has benefitted from remarks by an unknown referee and by the critical readings by G. Cicogna and S. Walcher, whom I warmly thank.

\end{document}